# Wire Resonator as a Broadband Huygens Superscatterer

Dmytro Vovchuk[1,2,*,=], Sergei Kosulnikov[1,3,=], Roman E. Noskov[1,=], and Pavel Ginzburg[1,4]

[1]Tel Aviv University, Ramat Aviv, Tel Aviv, 69978, Israel
[2]Department of Radio Engineering and Information Security, Yuriy Fedkovych Chernivtsi National University, Chernivtsi, 58012, Ukraine
[3]Department of Physics and Engineering, ITMO University, Saint Petersburg, 197101, Russia
[4]Center for Photonics and 2D Materials, Moscow Institute of Physics and Technology, Dolgoprudny, 141700 Russia

**Abstract:**

Interference phenomena render tailoring propagation of electromagnetic waves by controlling phases of several scattering channels. Huygens element, being a representative example of this approach, allows enhancement of the scattering from an object in a forward direction, while the reflection is suppressed. However, a typical resonant realization of Huygens element employs constructive interference between electric and magnetic dipolar resonances that makes it relatively narrowband. Here we develop the concept of a broadband resonant Huygens element, based on a circular array of vertically aligned metal wires. Accurate management of multipole interference in an electrically small structure results in directional scattering over a large bandwidth, acceding 10% of the carrier frequency. Being constructed from non-magnetic materials, this structure demonstrates a strong magnetic response appearing in dominating magnetic multipoles over electric counterparts. Moreover, we predict and observe very high-order magnetic hexadecapole (M16-pole) and magnetic triakontadipole (M32-pole) with quality factors, approaching 6,000. The experimental demonstration is performed at the low GHz spectral range. Our findings shed light on a simple approach for engineering compact and open electromagnetic devices (antennas, directional reflectors, refractors, etc.) able to tailor wave propagation in a broadband domain, concentrate strong magnetic field, and generate high-order magnetic multipoles.

*corresponding author dimavovchuk@gmail.com

[=] equal contribution

**Introduction**

Increasing scattering efficiencies of subwavelength structures is a long-standing objective of applied electromagnetic theory. Quite a few fundamental bounds have been proposed and subsequently challenged with advanced designs. For example, the celebrated Chu-Harrington limit defines minimally achievable quality factors (Q-factors) of subwavelength lossless resonators [1]. On another hand, it is relatively well known that the maximal scattering efficiency of a resonant subwavelength object does not depend on its size if internal material losses are neglected. In a single resonance case, a fundamental limit of scattering is $(2l' + 1)\lambda^2/(2\pi)$, where $\lambda$ is the free space wavelength and $l'$ is related to orbital angular momentum of a multipolar resonance ($l' = 1$ is the electric or magnetic dipolar case. Strictly speaking, the limit applies on geometries with a spherical symmetry). A vast majority of reported designs operates at dipolar resonances. In the case of lossless structures the upper 'dipolar' bound is $3\lambda^2/(2\pi)$ [2–6].

While subwavelength structures can be quite efficient scatterers, the penalty of size reduction is a dramatic growth in the Q-factor, described in a single resonance case by the Chu-Harrington bound. A strategy to bypath this limitation is to involve several spectrally overlapping multipolar resonances. This approach allows increasing the bandwidth while keeping strong scattering efficiency with help of cascading resonances. The concept of multipolar spectral overlap has been developed in the field of electrically small antennas in order to increase directivity. Specifically, the commonly accepted directivity bound is related to the antenna aperture ($A$) in units of the operational wavelength ($\lambda$) square as $G_{max} = 4\pi A/\lambda^2$, so that so-called Einstein's needle radiation (extremely high directivity) can be obtained in a deeply subwavelength structure if a large number of multipoles interferes constructively [7]. Quite a few theoretical approaches to achieve high directivity and superscattering have been reported. For example, core-shell spherical or cylindrical geometries were extensively explored. Those structures, having closed form analytical solutions (i.e. Mie theory), are subject to fast optimization and, hence, were found to promise a significant overcome of classical bounds [8–12].

However, practical limitations significantly degrade the theoretical predictions [13]. In particular, employment of high-order multipoles in subwavelength structures increases the stored near-field energy within a structure, giving rise to strong internal material losses. Another important issue is the fabrication tolerance. Any imperfection in respect to an original design shifts phases of multipoles and destroy their constructive interference. As an intermediate summary here, suggesting partial solutions to the beforehand mentioned challenges, is to avoid using lossy materials in the designs (e.g. PCB substrates) and employ geometrically simple elements to facilitate fabrication tolerances. Following these guidelines, we propose a set of straight copper wires within a styrofoam host as a subwavelength superscatterer.

Another topic, closely related to the scattering management, is the emerging field of metasurfaces [14–23] which allow tailoring electromagnetic energy flow with subwavelength thin patterned layers. In a vast majority of the reported designs, resonant elements are organized within two-dimensional arrays, and collective interaction phenomena are obtained. One of them is the backscattering cancellation based on the interference of electric and magnetic dipoles in a single particle. This effect was first proposed by Kerker et al. [24], and since then the concept has been advanced and generalized in many realizations, e.g.

[25–31]. Particles, which demonstrate a suppressed backward scattering, are also referred to as 'Huygens elements', as they might act as sources of secondary waves in Huygens–Fresnel diffraction theory. Being an interference phenomenon, *resonant* Huygens elements are inherently narrowband in a sharp contrast to a non-resonant case when electric and magnetic dipoles can constructively interfere in a broadband domain but with very low scattering efficiency [30,32].

Here we develop and experimentally demonstrate a new concept of a *broadband resonant* Huygens element, implemented as a wire resonator. It is based on a circular array of wires, embedded in a styrofoam host. The resonator supports a variety of multipolar resonances, which emerge from a near-field coupling between short metal wires. First, we perform numerical analysis of the scattering efficiency as a function of the number of wires and find the optimal number of wires meeting a trade-off between simplicity and performance of the structure. The eigenmodes of the optimized system are characterized by the numerical simulations, the Bloch theorem, and near-field measurements. Specifically, we predict and observe very high-order magnetic hexadecapole (M16-pole) and magnetic triakontadipole (M32-pole) with quality factors, approaching 6,000. Next, we study scattering performance of the structure by measuring and calculating scattering patterns and the scattering efficiency. Field decomposition with respect to Cartesian multipoles reveals contributions of various electric and magnetic multipoles to the eigenmodes and the scattering efficiency. We show that multipole interference leads to directional scattering over a large bandwidth, acceding 10% of the carrier frequency, and magnetic multipoles dominate over electric counterparts for almost all eigenmodes except for the lowest one. The excellent agreement between experiment, theory and numerical simulations is demonstrated.

The concept of a circular wire resonator should not be confused with circular antenna arrays, e.g. [33,34]. Near-field coupling in those electrically large structures is negligible and does not govern the response, while this type of interaction is the key for obtaining the broadband unidirectional scattering by a subwavelength element.

**The structure optimization**

An array of near-field coupled vertical wires supports a variety of multipolar modes. In a short-wire approximation and taking into account the mode degeneracy, the number of eigenmodes equals to ceil($N/2$) [33,35], where $N$ is number of wires. Hence, it is quite appealing to increase the number of wires within an array to achieve a richer mode structure. Nevertheless, strong near-field coupling in the subwavelength geometry sets certain limitations and complexity, which manifest itself in mode hybridization. Here we will explore a circular array of short metal wires, equidistantly distributed on a cylindrical surface (Fig. 1(a)). The first objective of the investigation aims on achieving a large scattering efficiency in an electrically small structure that will surpass the Chu-Harrington limit. For this purpose, a numerical optimization, based on the finite element method is performed. The investigated geometry consists of a variable number of half-wavelength thin metal wires (perfect electric conductor is taken of the initial analysis), alighted on a cylinder with the radius R = 33 mm. The wire length is chosen to be 60.7 mm, corresponding to a dipole wire resonance around 2 GHz, which is the relevant frequency range from wireless communications perspectives. The structure is illuminated with a plane wave, propagating along $Z$-axis and polarized along $Y$-axis.

Figs. 1(b and c) summarize the results, showing the scattering efficiency as a function of system's parameters. It is quite remarkable that the maximal efficiency approaches the value of 650 cm$^2$, which prevail the single channel limit (at 2 GHz) by a factor of 6. This improvement, however, saturates fast and can be obtained with an array of 5 wires (Fig. 1(c)). At the same time, the frequency, where the maximal scattering is obtained, weakly depends on the number of wires and stays close to a dipole resonance of a single wire. As it will be shown hereinafter, mode hybridization (bonding and antibonding [36]) causes splitting around an initial resonant frequency.

In order to provide a broadband high scattering efficiency and a rich mode family while keeping the relative simplicity of a circular wire resonator, we take an 11-wire structure (Fig. 1(c)) for further considerations.

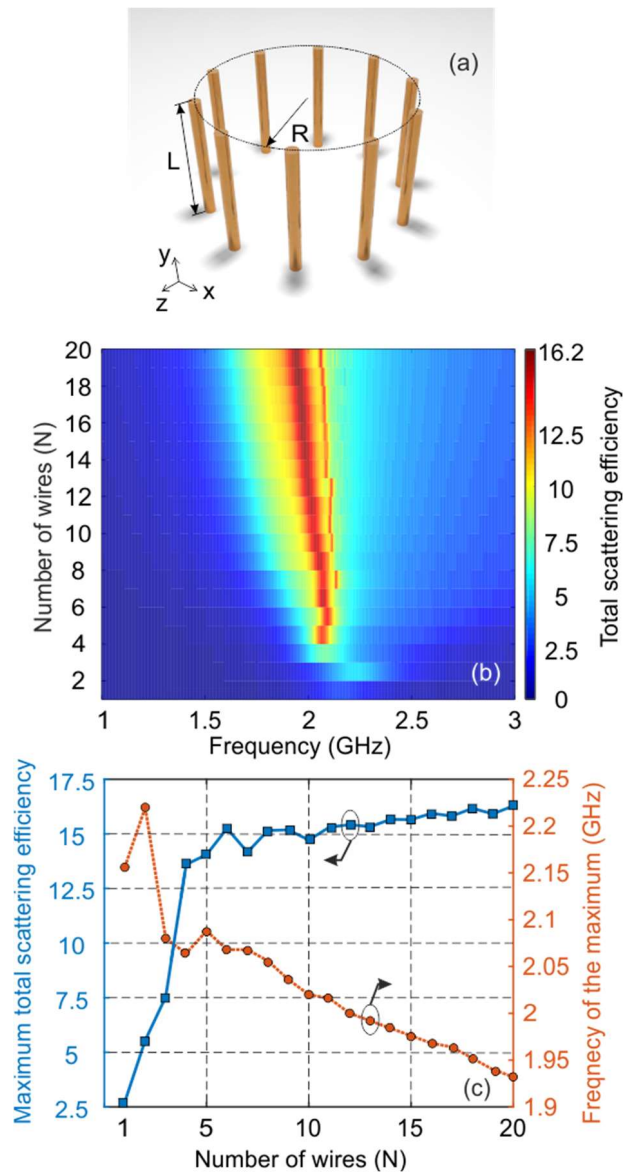

**Figure 1:** (a) Schematics of a circular wire array – a set of vertically aligned wires are equidistantly distributed on a cylinder's surface. (b), (c) Numerical analysis of the scattering efficiency (the scattering efficiency normalized by the geometric cross-section of the structure, perpendicular to the wave propagation direction (2R·L)) for a plane electromagnetic wave, propagating along Z-axis and polarized along Y-axis. (b) The total scattering efficiency as a function of frequency and the number of wires within the array. (c) The maximum scattering efficiency (left) and the corresponding frequency (right) as functions of the number of wires ($N$).

**Eigenmodes of a circular wire resonator**

The optimized structure of 11 vertically aligned copper wires, equidistantly distributed over a cylindrical surface (Fig. 2a), will be investigated next. Experimental conditions correspond to the following values – wires' diameter and length are 1 mm and 60.7 mm respectively, while the radius of the foam cylinder is 33 mm. Styrofoam, transparent for GHz electromagnetic radiation, has been used to host metal wires. Both numerical and experimental investigations to reveal electromagnetic properties of the structure include near-to-near (Fig. 2a), near-to-far (Fig. 2b) and far-to-far (Fig. 2c) field excitation and measurement methods. The near-field excitation (Tx probe) and the receiving/scanning (Rx probe) are realized as magnetic field loop probes, located near the sample surface. Setups for the far-field measurements include Tx and Rx horn antennas, allocated at the distance of 2.5 m from the sample to satisfy the Fraunhofer condition. The first configuration is needed to excite the entire modes of the structure (including dark weakly radiative ones) and to measure their field distributions. The second configuration detects the radiation efficiencies of the modes, while the last layout is the typical scattering scenario to test Huygens element performances.

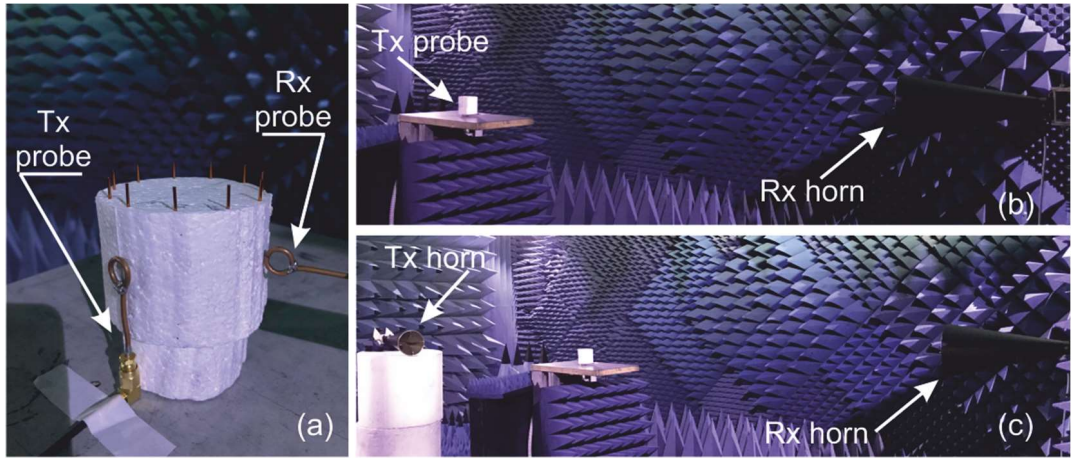

**Figure 2:** Photographs of the realized experimental setups in an anechoic chamber with the sample consisting of $N = 11$ copper wires equidistantly distributed on a styrofoam cylinder's surface for three considered types of investigations: (a) - near-field-to-near-field excitation (both Tx and Rx are realized with near-field magnetic loop probes); (b) - near-field-to-far-field excitation (Tx is a near-field magnetic loop probe and Rx is a far-field horn); (c) - far-field-to-far-field excitation (both Tx and Rx are realized as the far-field horn antennas).

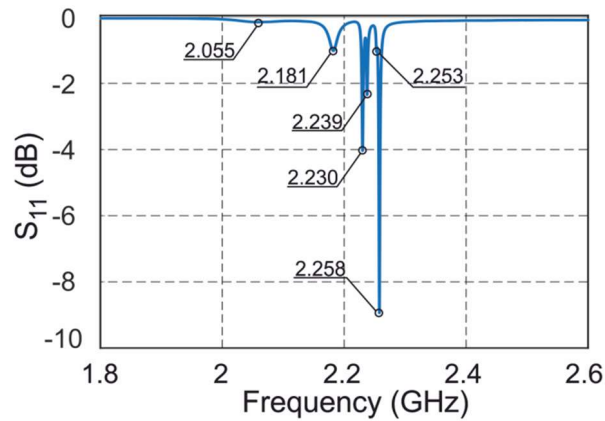

**Figure 3:** The measured reflection coefficient (abs($S_{11}$) in dB) spectrum. The excitation configuration appears in Fig. 2(a).

In order to probe the entire modes of the structure (at the expected frequency range - 1.8-2.6 GHz), the array is excited with a small non-resonant near field loop (Tx) and the complex reflection coefficient ($S_{11}$-parameter) spectrum is acquired. An optimal (most efficient excitation of maximal number of modes) position of Tx with respect to the structure is found to be at the middle of one of the wires. $S_{11}$-parameters spectra are presented in Fig. 3, which clearly show 6 high quality (high-Q) resonances.

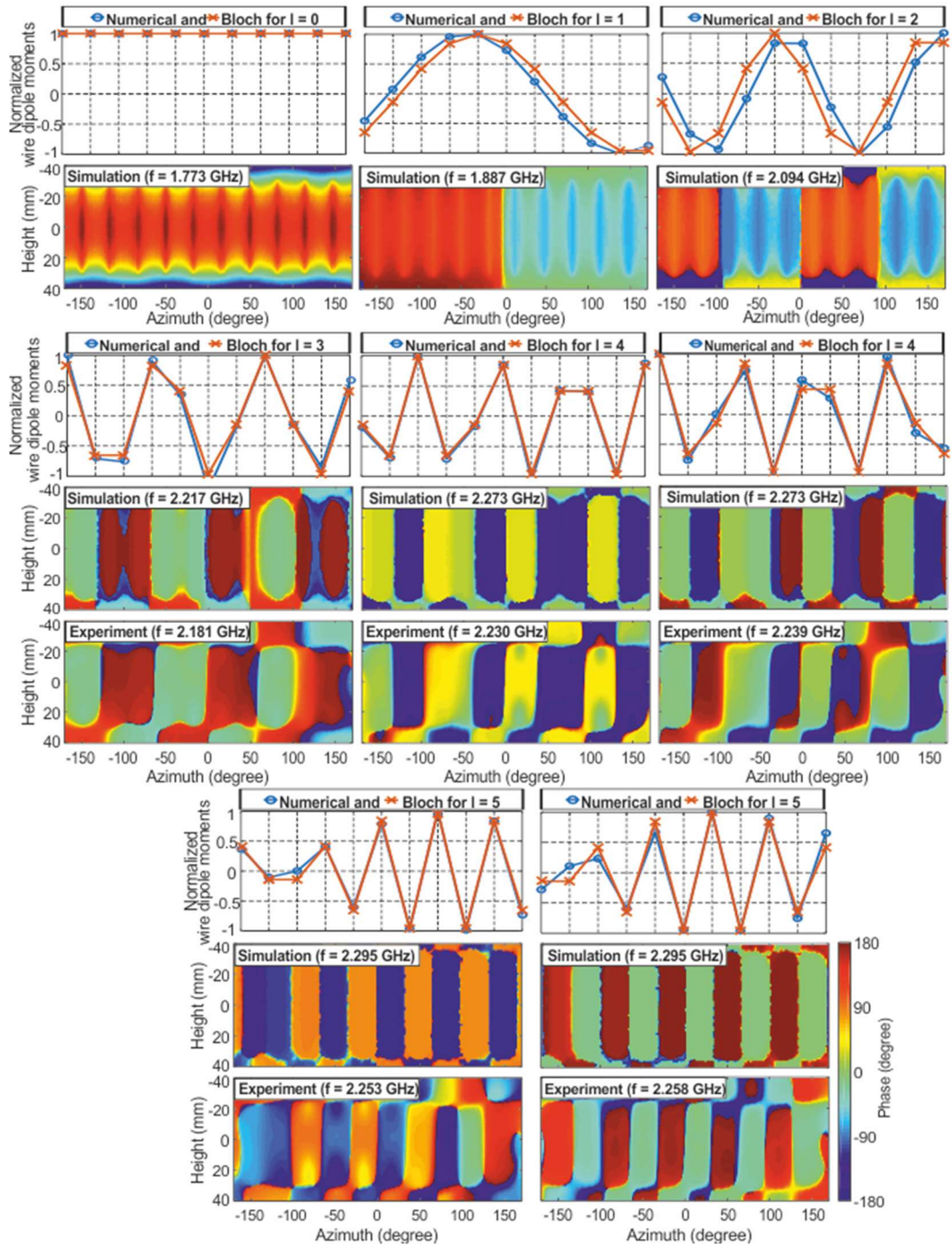

**Figure 4:** Characterization of eigenmodes within a circular wire array by the Bloch theorem, numerical simulations and near-field measurements. The numerical data is obtained from the eigenmode analysis, and experimental acquisition is performed with the near-field scan of azimuthal component of the magnetic field distributions (Fig. 2a). Resonances of the modes and their classification appear in Table 1. The eigenmodes with $l = 0$ and $l = 1$ have not been revealed in the experiment due to low Q-factors. The mode with $l = 2$ appears in $S_{11}$ but we do not provide for it the experimental phase distribution.

Next, we scan near-field distributions at the relevant frequencies corresponding to the minima of $S_{11}$-parameter (Figure 4). Magnetic field probe (a loop) is placed at the distance of 2 mm from the cylinder and moved around the structure (Fig. 2(a)). This circular scan is repeated for a different height of the probe, which maps the entire structure from the bottom to the top. The orientation of the probe allows detecting the azimuthal (within a good approximation) component of the magnetic field, including amplitude and phase. MiDAS system (ORBIT/FR Engineering Ltd.) is used to perform the experiment, while the sample is placed at the center of a rotating stage. Azimuthal locations of the wires correspond to the following angles: ±18°; ±54°; ±90°; ±126° and ±162°. An additional wire is situated at 180°, where Tx excitation probe is placed. The field scanning in the experiment is performed from -170° to +170°. Scanning along the vertical axis is executed in the range -40…+40 mm to cover the entire geometry.

Eigenmodes of a circular array of omni-directional scatterers (in plane in our case) can be naturally characterized by an azimuthal order, similarly to so called phase-modes in circular antenna arrays [33,34]. In accord to the Bloch theorem, phase variations of wire dipole moments obey geometrical periodicity of the structure as $p_n \sim \cos(2\pi \frac{n}{N} l)$, where $l$ is the mode azimuthal order and $n$ is the wire's ordinal number. Specifically, the 11-wire array possesses 6 eigenmodes.

| | *Experimental data* | | | | | | | |
|---|---|---|---|---|---|---|---|---|
| Azimuthal order | 0 | 1 | 2 | 3 | 4 | 4 | 5 | 5 |
| Dominating multipole | ED | MD | MQ | MO | M16-pole | M16-pole | M32-pole | M32-pole |
| $f$, GHz | - | - | 2.055 | 2.181 | 2.230 | 2.239 | 2.253 | 2.258 |
| $Q$-factor | - | - | 32.5 | 220 | 1316 | 1595 | 1211 | 2579 |
| | *Numerical data* | | | | | | | |
| $f$, GHz | 1.773 | 1.887 | 2.094 | 2.217 | 2.273 | | 2.295 | |
| $Q$-factor | 1.1 | 3.8 | 20 | 120 | 873 | | 5599 | |

**Table 1:** Characterization of the resonant modes, experimental and numerical data. Modes are characterized by their azimuthal order, dominating multipole, resonant frequency, and quality factor. ED and MD denote electric and magnetic dipoles, MQ – magnetic quadrupole, MO – magnetic octupole, M16-pole – magnetic hexadecapole, M32-pole – magnetic triakontadipole.

Figure 4 and Table 1 summarize the results of the numerical eigenmode analysis, predictions of the Bloch theorem, and the measurements. We underline excellent agreement between all these approaches. It is instructive to relate the classification by the azimuthal order to Cartesian multipoles [37]. In the lowest order mode ($l = 0$) there is no phase shift between wires that results in a dominating electric dipole (ED) response of the whole structure associated with strong scattering/radiation losses and a low Q-factor. An increase in the azimuthal order gives rise to alternating polarity of currents, exited on wires, and nontrivially configured loops, which serve as sources of magnetic field, so that $l = 1$ corresponds to magnetic dipole (MD), $l = 2$ – magnetic quadrupole (MQ), $l = 3$ – magnetic octupole (MO), $l = 4$ – magnetic hexadecapole (M16-pole), $l = 5$ – magnetic triakontadipole (M32-pole) (rigorous field decomposition with respect to Cartesian multipoles is presented in Fig. 6). The similar scenario has also been observed in circular arrays of plasmonic particles, where collective plasmon resonances give rise to artificial magnetism at optical frequencies [38,39]. Importantly, the higher azimuthal order also leads to the higher Q-factor as a result of stronger near field concentration and poorer scattering/radiation efficiency. In absence of Ohmic losses (or when those are negligible), scattering/radiation efficiency governs Q-factors which are remarkably high for an open resonator in cases of $l = 4$ and $l = 5$.

The resonant frequencies and Q-factors obtained in the experiment are fitted quite well with the numerical modelling. Accurate retrieval of sharp resonances is quite challenging and, hence, numerical predictions deviate from the experiment, especially in the case of high-order multipoles (Table 1). Experimental Q-factors are extracted from the $S_{11}$ spectrum, while the theoretical values are obtained from the eigenmode analysis. ED and MD modes, which are theoretically predicted, were not found in the experiment because of the relatively low Q-factor (below 20). In this case, resonant peaks are speared with tails of other nearby resonances.

Also, we note that the geometry of the real structure is not ideally symmetric. As a result, the modal degeneracy is removed, and the measurements show two couples of modes with $l = 4$ and $l = 5$ with slightly shifted resonant frequencies, while the Bloch theorem predicts a single mode per a single azimuthal order. This experimental aspect reflects the asymmetry in the realization and underlines the impact of fabrication tolerance on performances of highly resonant structures, as discussed in the introduction.

**Far-field signatures of high-order multipoles**

The next step is to estimate far-field signatures of the eigenmodes. It is worth noting that balancing those contributions will allow achieving broadband Huygens element, which will be discussed in the next section.

The experimental setup for far-field estimation is presented in Fig. 2(b). Here the sample is excited by the Tx near-field probe and the radiated patterns are acquired with the help of Rx horn antenna, located at the far-field region. Circular scan around the sample allows measuring the radiation in plane. Figure 5 shows far-field patterns (E-field) along with the overall in plane radiated power spectra. Signatures of the eigenmodes can be clearly identified. For example, 4-lobes E-field structure (at 2.055GHz) corresponds to the MQ eigenmode with $l = 2$. Other modes can be identified in a similar fashion, comparing Fig. 5 with Table 1. Higher order modes have lower radiation efficiencies and, as a result, the far-field patterns are noisier (data in the insets to Fig. 5 is normalized to unity for each one independently).

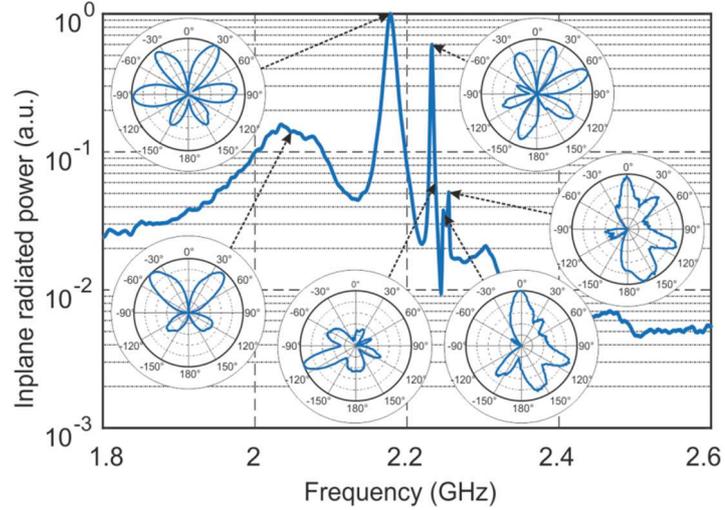

**Figure 5:** In-plane radiated power as a function of frequency. Insets - radiation patterns (E-field), the setup from Fig. 2(b). Frequencies correspond to resonant modes, which appear in Table 1. Radiation patterns are normalized to the maxima.

**Analysis of the scattering efficiency**

Next, we investigate the performance of the structure, operating as a scatterer in free space (the configuration in Fig. 2(c)). Optical theorem is used to evaluate the total scattering efficiency from the imaginary part of the forward one in the case of the experimental acquisition [40,41]. Experimental and numerical results are summarized in Fig. 6 (a). While the experimental data is colored by an additional oscillatory patter, which is quite typical to experiments, performed in an anechoic chamber (e.g. [42,43]), a good correspondence with the numerical data can be observed. Main high-Q peaks are clearly visible, and their spectral position is almost perfectly predicted by the numerical modelling (40 MHz shift was manually compensated to obtain more accurate layout of the results). It is also worth noting the similarity between Figs. 5 and 6(a), despite different experimental layouts, used to obtain the data. This is the manifestation of strong resonant behavior, which predominates other mechanisms (e.g. absorption) that might play a role. High-Q modes are the preferable channels for radiation and scattering; hence, they are visible in both cases (Figs. 5 and 6).

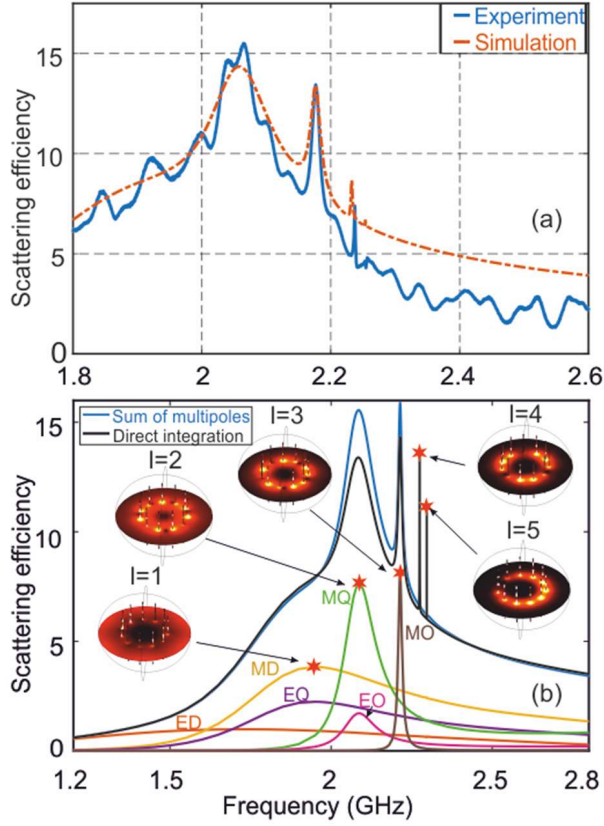

**Figure 6:** Analysis of the scattering efficiency spectrum. (a) The comparison of measured and simulated scattering efficiencies. (b) Cartesian multipole decomposition of the scattering efficiency. Multipole contributions are marked with colors: ED and MD – electric and magnetic dipoles; EQ and MQ – electric and magnetic quadrupoles; EO and MO – electric and magnetic octupoles. Blue line denotes the summation of the multipolar contributions, black line marks scattering efficiency derived by the direct far-field integration. Insets show the near-field distribution of the magnetic field in the equatorial plane and current distribution on the wires (white cones) at the eigenmode frequencies. Both figures correspond to the experimental layout in Fig. 2(c).

In order to reveal the contribution of eigenmodes to the scattering performance, we perform the decomposition of the simulated scattered field with respect to Cartesian multipoles with 3 orders included as follows [37]:

$$E_{scat\_i} = \frac{k^2}{4\pi\varepsilon_0 r} e^{ikr} \left[ (n_i n_j - n^2 \delta_{ij}) \left( -\left( p_j + \frac{ik}{c} T_j^{(e)} + \frac{ik^3}{c} T_j^{(2e)} \right) + \frac{ik}{2} \left( \bar{\bar{Q}}_{jk}^{(e)} + \frac{ik}{c} \bar{\bar{T}}_{jk}^{(Qe)} \right) n_k \right. \right.$$
$$\left. + \frac{k^2}{6} \left( \bar{\bar{O}}_{jkp}^{(e)} + \frac{ik}{c} \bar{\bar{T}}_{jkp}^{(Oe)} \right) n_k n_p \right) + \varepsilon_{ikj} n_k \left( -\frac{1}{c} \left( m_j + \frac{ik}{c} T_j^{(m)} \right) + \frac{ik}{2c} \left( \bar{\bar{Q}}_{jp}^{(m)} + \frac{ik}{c} \bar{\bar{T}}_{jp}^{(Qm)} \right) n_p + \frac{k^2}{6c} \bar{\bar{O}}_{jpl}^{(m)} n_p n_l \right) \right].$$

We assume $\exp(-i\omega t)$ time-dependence and Einstein's summation notation, where vectors and tensors are denoted by lower-case indexes. The following parameters are defined: $k$ is the vacuum wave-number, $r = |\mathbf{r}|$ the distance between the scatterer's center and the observation point, $\varepsilon_0$ the vacuum permittivity, $n_i = \mathbf{r}/r$ ; $p_j$ and $m_j$ basic electric dipole (ED) and magnetic dipole (MD), $\bar{\bar{Q}}^{(e)}_{jk}$ and $\bar{\bar{Q}}^{(m)}_{jp}$ electric and magnetic quadrupoles (EQ and MQ), $\bar{\bar{O}}^{(e)}_{jkp}$ and $\bar{\bar{O}}^{(m)}_{jpl}$ electric and magnetic octupoles (EO and MO). The toroidal moments are marked with *T*. Superscripts in round brackets indicate their belonging to the corresponding basic multipole moments. Explicit expressions for multipole tensors can be found in Ref. [37]. The differential scattering efficiency is given by

$$dP_{scat} = \frac{1}{2}\sqrt{\frac{\varepsilon_0}{\mu_0}}|\mathbf{E}_{scat}|^2 d\Omega.$$

Normalization of $P_{scat}$ to the incident wave energy flux and the geometrical efficiency of the scatterer leads to the following expression for the scattering efficiency:

$$\sigma_{scat} = \frac{2}{2RL}\sqrt{\frac{\mu_0}{\varepsilon_0}}\frac{P_{scat}}{|\mathbf{E}_0|^2},$$

where $|\mathbf{E}_0|$ is the amplitude of the incident field.

Figure 6(b) summarizes the results. The validity of the decomposition is evidenced by the agreement between the sum of multipole contributions and the scattering efficiency obtained by the integration of scattered field. The only significant differences appear for the highest order modes with $l = 4$ and $l = 5$ because the order of multipoles corresponding to them is beyond the set of 6 multipoles considered. However, accounting for the symmetry of these modes, one can assume that the dominating multipoles for them are M16- and M32-poles, respectively.

It is important to note that magnetic multipoles prevail over electric counterparts for all eigenmodes except for the lowest order ED mode. This is the result of alternating wire polarizations serving as sources of magnetic field which is shown in the insets of Fig. 6(b) and confirmed by the phase patterns, presented in Fig. 4.

**Broadband Huygens element**

Finally, we demonstrate a broadband Huygens element operation, which relies on the multipolar resonant interference. For this purpose, the scattering efficiencies in the forward and backward directions with respect to the wave incidence should be compared, and regions where the first one predominates the second should be found. The forward and the backward scattering efficiencies as well as their ratio are shown in Figure 7, demonstrating both experimental and numerical results.

Once the forward-to-backward ratio prevails the value of 2 (rather arbitrary yet reasonable criteria), the element can be considered as Huygens. One can observe a minimum of the backward scattering at 1.93 GHz in the experiment (Fig. 7(a)) and at 1.6 GHz in the simulation (Fig. 7(b)) which corresponds to the maximum in the forward-to-backward ratio. This is the result of the classical Kerker effect [24] associated with equal and in-phase ED and MD contributions to the scattering (Fig. 6(b)). At higher frequencies, several mixed predominantly magnetic multipoles contribute to the scattering. Although this mixture is changing with the frequency, the main contribution to broadband directional scattering is provided by low-Q MD and MQ modes. As a result of constructive multipole interference also known as the generalized Kerker effect, the forward scattering is always dominating. Hence, the directional resonant scattering is achieved over a relatively broad spectral range, covering 10s of percent of the bandwidth (Figs. 7(c) and (d) for the experiment and the theory). Scattering patterns, which appear as insets in Fig. 7(d), clearly show dominating forward scattering. It should be noted that the effect here is different from other broadband Huygens elements, based on nonresonant elements [30,32]. In nonresonant cases the scattering efficiencies are much weaker and spectral control over the effect is more challenging if even possible.

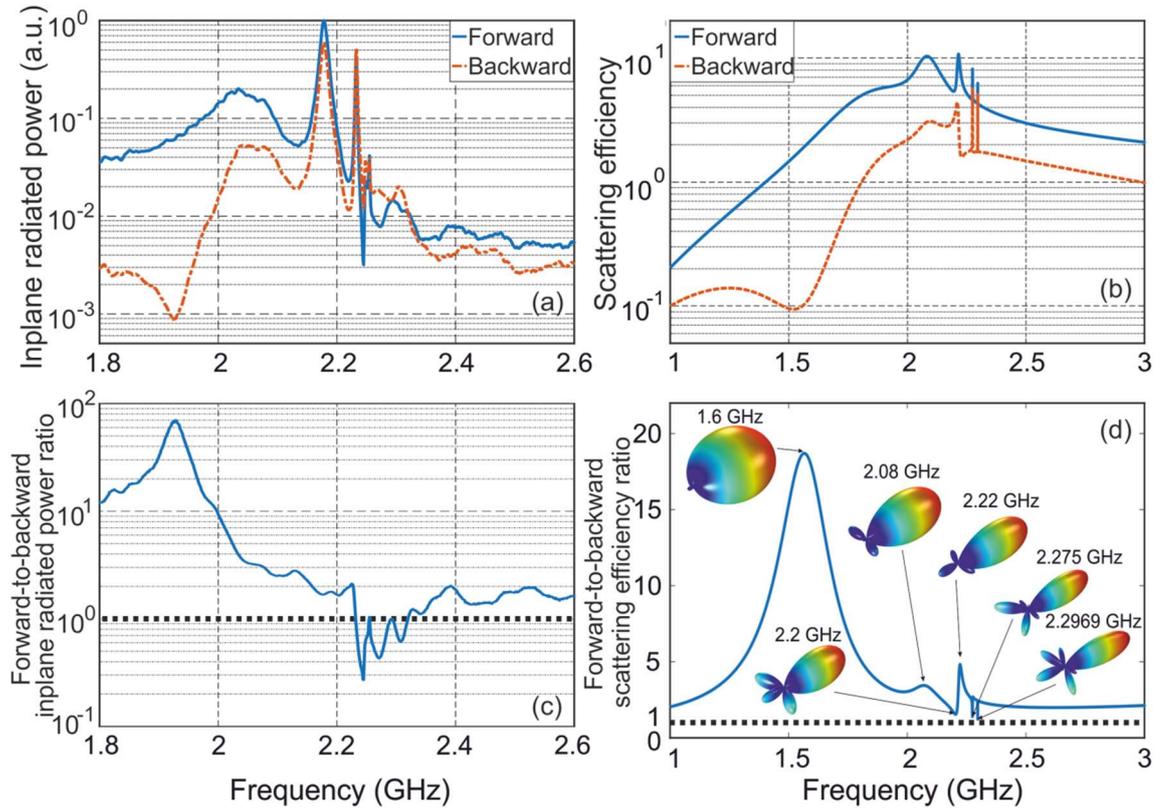

**Figure 7:** Forward and the backward scattering efficiency spectra: (a) – experimental, (b) - numerical. Forward-to-backward scattering ratio: (c) - experimental, (d) - numerical. Insets show far-field scattering diagrams. Measurements have been performed for two directions only, while in simulations the forward and backward scattering efficiencies have been calculated by integrations of the far-field over half-spheres.

**Conclusion and Outlook**

The concept of a broadband resonant Huygens element based on a circular array of vertically aligned metal wires has been proposed and experimentally demonstrated. The effect stems from spectral cascading a variety of system's eigenmodes appearing as a result of a strong near-field coupling between dipolar responses of individual wires. Field expansion with respect to Cartesian multipoles shows predominant contribution of magnetic multipoles with respect to electric ones for almost all eigenmodes except from the lowest one. Very high-order magnetic hexadecapole (M16-pole) and magnetic triakontadipole (M32-pole) with quality factors, approaching 6,000, have been predicted and mapped with a near-field scanner. Multipoles contribution to the Huygens element performance have been shown.

Scattering efficiency of the demonstrated configuration is 15 times larger than its geometrical cross section and, at the same time, the device demonstrates broadband forward scattering capabilities, obtained at a bandwidth ~ 10% of the carrier frequency in the GHz range.

Resonant Huygens elements of this kind can find a use in new frequency-selective surfaces designs, antenna isolation devices, directional reflectors, refractors, and many others. It is also worth noting that the concept can be scaled in frequency and the resonant cascading approach can be mapped on the optical domain. Here light harvesting is an important objective, contributing to green energy efforts and flat miniaturized imaging optics to name just few.

**Acknowledgments**

The research was supported in part by Pazy foundation, ERC StG 'In Motion', and Ministry of Science and Technology (project "Integrated 2D&3D Functional Printing of Batteries with Metamaterials and Antennas".